\documentclass{jfm}

\usepackage{graphicx}
\usepackage{natbib}

\ifCUPmtlplainloaded \else
  \checkfont{eurm10}
  \iffontfound
    \IfFileExists{upmath.sty}
      {\typeout{^^JFound AMS Euler Roman fonts on the system,
                   using the 'upmath' package.^^J}%
       \usepackage{upmath}}
      {\typeout{^^JFound AMS Euler Roman fonts on the system, but you
                   dont seem to have the}%
       \typeout{'upmath' package installed. JFM.cls can take advantage
                 of these fonts,^^Jif you use 'upmath' package.^^J}%
      }
  \else
  \fi
\fi

\ifCUPmtlplainloaded \else
  \checkfont{msam10}
  \iffontfound
    \IfFileExists{amssymb.sty}
      {\typeout{^^JFound AMS Symbol fonts on the system, using the
                'amssymb' package.^^J}%
       \usepackage{amssymb}%
         \let\leq=\leqslant
         \let\geq=\geqslant
      }{}
  \fi
\fi

\ifCUPmtlplainloaded \else
  \IfFileExists{amsbsy.sty}
    {\typeout{^^JFound the 'amsbsy' package on the system, using it.^^J}%
     \usepackage{amsbsy}}
    {}
\fi

\newsavebox{\astrutbox}
\sbox{\astrutbox}{\rule[-5pt]{0pt}{20pt}}

\newcommand\p{\ensuremath{\partial}}

\def\d{\:{\rm d}}

\usepackage{color}

\def\ba#1#2\ea{\begin{align}\label{#1}#2\end{align}}
\def\bsa#1#2\esa{\begin{subequations}\label{#1} \begin{align}#2\end{align} \end{subequations}}

\def\L{\mbox{------}}
\def\dashL{\mbox{-~-~-}}
\def\dotL{\mbox{$\cdot\ \cdot\ \cdot$}}

\def\cdotL{\mbox{--- $\cdot$ ---}}

\def\dashdot{\mbox{- $\cdot$ -}}

\def\d{\textrm{d}}

\def\p{\partial}
\def\be{\begin{eqnarray}}
\def\ee{\end{eqnarray}}
\def\bes{\begin{subeqnarray}}
\def\ees{\end{subeqnarray}}
\def\f{\frac}
\def\lp{\left(}
\def\rp{\right)}
\def\lb{\left[}
\def\rb{\right]}
\def\lcb{\left\{}
\def\rcb{\right\}}

\def\z{\zeta}

\def\befi{\begin{figure}}
\def\eefi{\end{figure}}
\def\theequation{\thesection.\arabic{equation}}

\def\bce{\begin{center}}
\def\ece{\end{center}} 
\def\nn{\nonumber} 
\def\L{\mbox{------}}
\def\dashL{\mbox{-~-~-}}
\def\dotL{\mbox{$\cdot\ \cdot\ \cdot$}}

\def\cdotL{\mbox{-- $\cdot$ --}}

\usepackage{color}
\definecolor{orange}{RGB}{255,127,0}

\def\N{\mathcal N}

\def\red{\textcolor{red}}
\def\blue{\textcolor{blue}}
\definecolor{darkblue}{RGB}{83,0,93}
\usepackage{graphicx,epsfig,epstopdf}\usepackage{soul}\def\co{{\mathcal O}}

\title{Landslide Tsunamis in Lakes}
\author [Couston, Mei \& Alam]%
{L\ls O\ls U\ls I\ls S\ls -\ls A\ls L\ls E\ls X\ls A\ls N\ls D\ls R\ls E\ns C\ls O\ls U\ls S\ls T\ls O\ls N,$^1$
C\ls H\ls I\ls A\ls N\ls G\ns C\ls .\ns M\ls E\ls I,$^2$ \and
M\ls O\ls H\ls A\ls M\ls M\ls A\ls D\ls -\ls R\ls E\ls Z\ls A\ns A\ls L\ls A\ls M,$^1$\break} \affiliation{$^1$ Department of Mechanical Engineering, University of California, Berkeley, CA 94720, USA\\
$^2$ Department of Civil and Environmental Engineering, Massachusetts Institute of Technology, Cambridge, MA, 02139, USA\\
[\affilskip]}

\usepackage{amsmath}

\begin{document}
\date{\today}
\maketitle

\begin{abstract}

 Landslides plunging into lakes and reservoirs can result in extreme wave runup at shores. This phenomenon has claimed lives and caused damage to near-shore properties. Landslide tsunamis in lakes are  different from  typical earthquake tsunamis in the open ocean in that  (i) the affected areas are usually within the near-field of the source, (ii) the highest runup occurs within the time period of the geophysical event, and (iii) the enclosed geometry of a lake does not let the tsunami energy escape. 
 To address the problem of transient landslide tsunami runup and to predict the resulting inundation, we utilize a nonlinear model equation in the Lagrangian frame of reference. The motivation for using such a scheme lies in the fact that the runup on an inclined boundary is directly and readily
computed in the Lagrangian framework without the need to resort to approximations.
 In this work, we investigate the inundation patterns due to landslide tsunamis in a lake.  We show by numerical computations that Airy's approximation of  an irrotational theory using Lagrangian coordinates  can  legitimately predict runup of large amplitude. We also demonstrate that in a lake of finite size  the highest runup  may be magnified by constructive interference between   edge-waves that are trapped along the shore and multiple  reflections of outgoing waves from opposite shores, and may  occur somewhat  later  after the first inundation.

\end{abstract}

\section{Introduction}

Landslide generated large waves are often reported in lakes, bays and large reservoirs. A tragic event of this type happened in 1963 when a massive landslide behind Vajont dam in Italy resulted in   large waves that overflowed the dam crest  and took more than 2000 lives in the neighboring villages \cite[e.g.][]{Genevois2005}. The overflow of water was estimated at 30 million cubic meters. Landslide tsunamis and associated damages have been reported for numerous lakes and partially-enclosed bodies of water such as Spirit lake \cite[in Washington State,][]{Voight1983}, Lake Loen \cite[in Norway,][]{Jorstad1968,Bryant2008}, Lake Tahoe \cite[California,][]{Gardner2000}, Lituya Bay \cite[Alaska, with $\sim$500m runup called Mega-Tsunami,][]{Fritz2009,Weiss2009}, and frequently at lake Roosevelt \cite[in Washington state,][]{Lockridge1990,Bryant2008} {and in the Volga river in Russia \cite[see][]{Didenkulova2007a}}. Aside from taking lives and causing damage to nearshore houses and facilities, lake tsunamis make navigation hazardous by moving mud, stones and other debris such as fallen trees and destroyed houses.

Along  an open coast the first few leading waves of a tsunami are usually responsible for the majority of the destruction. Tsunamis in lakes or in other enclosed water bodies are different in that (i) the affected areas are near the source of disturbance, (ii) destructive runups may occur close to the time of  the initial entry, and (iii) waves do not escape  or disperse but are reflected back and forth until damped into heat and turbulence. {Thus far numerical prediction of   runup on shores of complex   bathymetry   is still an  arduous task, due to a number of difficulties such as: (i)   moving shoreline, (ii) onset  and propagation of breaking waves, and (iii)    interaction  with stationary or moving objects. To overcome all these challenges, direct numerical simulation taking full  account of nonlinearity and turbulence is necessary. In practice, however, coastal planning and warning procedures 
require efficient forecasting schemes based on approximate models. 

For tsunamis originating from distant earthquakes, numerical models based on the linear theory can be quite adequate to account for dispersion and scattering during transoceanic propagation. Upon entering  the continental shelf,  wave amplitude  increases and frequency dispersion diminishes. While Boussinesq approximation is a good basis for modeling weakly nonlinear long waves before reaching the shore, coastal flooding calls for  mathematical models capable of  predicting highly nonlinear waves in very shallow water or on dry land.

Airy's approximation for nonlinear long waves has been used in the conventional Eulerian framework to predict the runup of moderately steep waves on beaches, with amplitudes comparable or larger than the local water depth. On this basis, an analytical theory for the runup on an infinitely long plane beach was put forward by \cite{Carrier1958}. The theory is based on the hodograph transformation of the shallow-water equations, and has the advantage of including the shoreline  motion in the final solution. While several interesting extensions have been investigated {\cite[e.g.][]{Tuck1972,Spielvogel1975,Tadepalli1994,Kanoglu2004,Zahibo2006,Didenkulova2007,Madsen2010a,Rybkin2014}}, the complexity of the solution and its restriction to idealized problems have limited its use \cite[see][]{Pelinovsky1992,Carrier2003}. To account for complex bathymetries and 3D problems, numerical codes have been developed by patching computations based on Airy's or Boussinesq's approximation near the shore and linearized approximation far offshore \cite[see for instance][]{Yeh1996}.

   Discrete computations based on Airy's equations in the Eulerian formulation require special care in the prediction of the moving shoreline. The challenge is that the horizontal extent of the computational domain with fixed grid points must be adjusted in time as waves run up and down. Different methods developed to predict the horizontal motion of the wet/dry interface can be found in the literature \cite[see for instance][]{Balzano1998,Lynett2002}. {These so-called wetting/drying algorithms are, however, approximate with specific accuracy/efficiency trade-offs, and their implementation is difficult to generalize as they typically depend on the specific model equations used (e.g. Boussinesq or Airy) as well as the particular numerical method chosen \cite[e.g. finite volume or finite element, c.f. e.g.][]{Tchamen1998,Medeiros2013}}. A convenient alternative is to work in the less-used Lagrangian frame of reference. In the Lagrangian framework the fluid flow is obtained by following the trajectory of each fluid particle $\vec x(\vec a,t)=(x,y,z)$, which is regarded as an unknown function of the fluid particle's initial  position $\vec a=(a,b,c)$ and the time $t$. The elevation and horizontal extent of the free surface are thus known at all times, as required by the kinematic condition. Therefore an immediate advantage of the Lagrangian formulation is that the free surface and the moving shoreline are known a priori and are defined by their initial positions.

	Historically, Airy was the first to derive the nonlinear  long-wave equation for a horizontal seabed using Lagrangian coordinates \cite[][]{Airy1841,Lamb1932}. 
The solution for the runup of small amplitude (linear) waves on a uniformly sloping beach was later obtained in the Lagrangian framework by \cite{Miche1944}, and was rederived in the limit of long waves by \cite{Shuto1967}. 
	Interestingly this linear runup solution is identical to the analytical prediction based on the nonlinear long-wave equation in Eulerian coordinates for periodic water waves \cite[c.f.][]{Carrier1958}. The Lagrangian theory for shallow-water waves was subsequently expanded by \cite{Shuto1968,Shuto1972,Shuto1978,Goto1979a,Goto1979b,Goto1980,Johnsgard1997b} and \cite{Fujima2007} to account for nonlinearity, arbitrary seabed topography and bottom deformation in two horizontal dimensions. Weakly nonlinear and weakly dispersive equations, similar to the  Boussinesq equations in Eulerian coordinates, have also been investigated in the Lagrangian framework \cite[c.f.][]{Pedersen1983,Zelt1986,Jensen2003}, with applications to wave runup investigations and harbor oscillations \cite []{Zelt1990}. 

In this article we shall use the long-wave approximation of Airy in Lagrangian framework \cite[][]{Johnsgard1997b,Fujima2007} for the numerical prediction of two dimensional tsunamis, created by landslides, in a shallow lake. We solve the governing equations with a fourth order Runge-Kutta method for time integration, and a compact finite-difference scheme for spatial differentiation. For the wave generation mechanism, we consider a solid subaerial slide moving at a constant speed down a sloping beach of an enclosed basin. Physical implications of three dimensionality on the wave pattern are discussed and quantitative improvements from linear approximation are assessed.

  
\section{Long-wave equations in Lagrangian coordinates}
Consider wave propagation on the surface of a homogeneous, inviscid and incompressible fluid. We define a Cartesian coordinate system with $x$- and $y$-axis on the mean free-surface and $z$-axis positive upward. Let   the initial and current locations of a fluid particle  be denoted respectively   by $(a,b,c)$ and $(x,y,z)$. Then the equation for mass conservation reads
\begin{equation}
\label{901}
\f{\p (x,y,z)}{\p (a,b,c)}=1,
\end{equation} 
and the momentum equation requires \cite[e.g.][]{Lamb1932}
\begin{equation}
\label{902}
\lp \begin{array}{ccc} x_a & y_a & z_a \\x_b & y_b & z_b \\x_c & y_c & z_c \end{array} \rp  \lcb \begin{array}{c} x_{tt} \\y_{tt} \\z_{tt}+g \end{array} \rcb+\f{1}{\rho}\lcb \begin{array}{c}p_a \\p_b \\p_c \end{array} \rcb=0.
\end{equation}
Assuming that the atmospheric pressure on  the free-surface ($c=0$) is uniform, the dynamic boundary conditions become
\bsa{903}
x_{tt}x_a+y_{tt}y_a+(z_{tt}+g)z_a=0,\hspace{1cm} c=0, \label{9031}\\
x_{tt}x_b+y_{tt}y_b+(z_{tt}+g)z_b=0,\hspace{1cm} c=0. \label{9032} 
\esa
Note that the free surface height above the still water level is given by $\eta=z(a,b,0,t)$. Assuming a time-varying water depth $h(x,y,t)$, the seabed is represented by $c=-h_0(a,b)=-h(a,b,t=0)$. Then the kinematic boundary condition on the bottom reads 
\be\label{904}
z(a,b,c,t)=-h(x,y,t),\hspace{1cm} c=-h_0(a,b).
\ee
As in Airy's theory in Eulerian coordinates we consider long waves in shallow water and  assume that the vertical displacement of fluid particles, i.e. the wave amplitude $A$, is comparable to the typical waterdepth $H$, but much smaller than the horizontal length scale $L$, i.e.
\be \label{9041} 
\mu= \f{H}{L}\ll 1,~~\f{A}{H}=\co(1). 
\ee
The relevant set of approximate equations for long-wave propagation will be obtained by employing dimensionless variables, denoted by asterisks, as follows
\be\label{852}
(x,y,a,b)=L(x^*,y^*,a^*,b^*),~~~(z,c,h_0,h,\eta)=H(z^*,c^*,h_0^*,h^*,\eta^*),~~~t=\f{L}{\sqrt{gH}}t^*. \hspace{0.6cm}
\ee
From this point on all our equations will be in the dimensionless form and we will drop the asterisks for notational simplicity. 
We define the fluid displacement vector $\vec X(a,b,c,t) = (X,Y,Z)$ by
\be \label{908}
x=a+X,~~~y=b+Y,~~~z=c+Z.
\ee
Expanding the displacement vector in a power series of the vertical coordinate, i.e.
\ba{912} 
\vec X = \sum_{n=0}^{\infty} c^n \vec X^{(n)}(a,b,t),
\ea
and invoking irrotationality, viz.,
\ba{irrot}
\f{\p(x,y,y_t)}{\p(a,b,c)} = \mu^2\f{\p (x,z_t,z)}{\p(a,b,c)}, \f{\p (x,y,x_t)}{\p(a,b,c)} = \mu^2\f{\p(z_t,y,z)}{\p(a,b,c)}, \f{\p(y_t,y,z)}{\p(a,b,c)} = \f{\p(x,x_t,z)}{\p(a,b,c)},
\ea
it can be shown that $(X,Y) = (X^{(0)},Y^{(0)})  + \co(\mu^2)$ and $Z = Z^{(0)} + cZ^{(1)} + \co(\mu^2)$. This implies, as expected for shallow-water waves, that the horizontal flow is vertically uniform, and as a result the pressure is hydrostatic. To the leading order, it is thus sufficient to solve the free surface conditions \eqref{903} for the variables $X^{(0)}$ and $Y^{(0)}$ only. Noting that ($X_0,Y_0$) are independent of $c$, after some algebra, from \eqref{903} we obtain 
\bsa{927}
&X^{(0)}_{tt}=\frac{Y^{(0)}_a Z^{(0)}_b - \lp 1+Y^{(0)}_b\rp Z^{(0)}_a}{1+X^{(0)}_a+Y^{(0)}_b+X^{(0)}_aY^{(0)}_b-X^{(0)}_bY^{(0)}_a},\label{9271}\\
&Y^{(0)}_{tt}=\frac{X^{(0)}_b Z^{(0)}_a - \lp 1+X^{(0)}_a\rp Z^{(0)}_b}{1+X^{(0)}_a+Y^{(0)}_b+X^{(0)}_aY^{(0)}_b-X^{(0)}_bY^{(0)}_a},\label{9272}
\esa
where
\ba{928}
Z^{(0)} = h_0(a,b)Z^{(1)} + h_0(a,b)-h(a+X^{(0)},b+Y^{(0)},t),
\ea
is the vertical displacement of the free surface as obtained from the  seabed condition \eqref{904}, 
with
\ba{929}
Z^{(1)}=-\frac{X^{(0)}_a+Y^{(0)}_b+X^{(0)}_aY^{(0)}_b-X^{(0)}_bY^{(0)}_a}{1+X^{(0)}_a+Y^{(0)}_b+X^{(0)}_aY^{(0)}_b-X^{(0)}_bY^{(0)}_a},
\ea
which is due to the leading order equation \eqref{901} for mass conservation. Equation \eqref{927} is dispersionless and is valid for simulation time of $t \leq \co(\f{1}{\mu})$.

Once the solution to the Lagrangian equations \eqref{927} is known, the free surface elevation is obtained from $\eta(x,y,t)=Z(a,b,c=0,t)$. To find the free surface elevation in the Eulerian framework, i.e. $\eta(x,y,t)$, we need to invert the nonlinear system
 \be x=a+X^{(0)}(a,b,t),~~y=b+Y^{(0)}(a,b,t),\ee  to get 
\ba{510}
a=a(x,y,t),~~~b=b(x,y,t),~~~c=0, 
\ea 
and substitute the results into \eqref{928}. 
This transformation can be done as long as the Jacobian, defined by $\f{\p (x,y)}{\p (a,b)}$, is nonzero \cite[see][for more details]{Zelt1986}.

 %
 \subsection{One dimensional limit}
For  one-dimensional propagation ($Y=\p/\p b=0$) the continuity equation \eqref{929} becomes
\ba{918}
Z^{(1)}=-\f{X^{(0)}_a}{1+X^{(0)}_a}.
\ea
With the help of \eqref{928}, the free-surface condition \eqref{9271} is then simplified to
\ba{919}
X^{(0)}_{tt}\lp1+X^{(0)}_a\rp+\lb h_0(a)-h(a+X^{(0)},t) \rb_a-\lb \f{h_0(a) X^{(0)}_a}{1+X^{(0)}_a}\rb_a =0. 
\ea
This equation will be solved numerically for model validation in section \S3.2. Note that in a case where the depth is constant everywhere, i.e.  $h=h_0=1$; \eqref{919} further reduces to
\ba{920}
X^{(0)}_{tt}=\f{ X^{(0)}_{aa}}{\lp 1+X_a^{(0)}\rp^3},
\ea
which was first given by Airy and is similar to the equation that governs the oscillations of nonlinear strings \cite[][]{Zabusky1962}.

%
\subsection{Linearized limit}\label{appb}
Assuming that $\vec X\sim\co(\epsilon)\ll1$ the linearized form of \eqref{929},\eqref{9271} and \eqref{9272} is obtained as 
\bsa{921}
& Z^{(1)}=-X^{(0)}_a-Y^{(0)}_b, \label{9211}\\
& X^{(0)}_{tt}=-Z^{(0)}_a, \label{9212}\\
& Y^{(0)}_{tt}=-Z^{(0)}_b. \label{9213}
\esa
%
Let us now consider that the seafloor deforms over time because of the passage of a landslide. Assuming this perturbation to be small we can rewrite the water depth as $h(x,y,t)= h_0(x,y)+f(x,y,t)$ where $f\sim \co(\epsilon)$. In this case, by Taylor expanding $h$ and substituting \eqref{9211} in \eqref{928}, the leading order free-surface height is obtained as
\ba{922}
\eta(x,y,t)=-f(x,y,t)-\nabla_{h}\lcb h_0 \lb \begin{array}{c} X^{(0)} \\ Y^{(0)} \end{array} \rb\rcb
\ea  
where $\nabla_{h}=(\p/\p a,\p/\p b)$. 
Taking the second time-derivative of \eqref{922}, and using \eqref{9212},\eqref{9213} we obtain
%
%
\ba{925}
\eta_{tt}+f_{tt}&=-h_0\lp X^{(0)}_{a,tt}+Y^{(0)}_{b,tt}\rp -h_{0,a}X^{(0)}_{tt}-h_{0,b}Y^{(0)}_{tt},\nn\\
&=h_0\lp Z^{(0)}_{aa}+Z^{(0)}_{bb}\rp +h_{0,a}Z^{(0)}_a+h_{0,b}Z^{(0)}_b,\nn\\
&= [h_0Z^{(0)}_a]_a+[h_0Z^{(0)}_b]_b .
\ea
Since $\eta=Z^{(0)}$, $X=x-a\sim\co(\epsilon)$ and $Y=y-b\sim\co(\epsilon)$, the above equation readily reduces to the classical shallow-water equation in the Eulerian frame of reference 
\ba{926}
\eta_{tt}+f_{tt}=[h_0(x,y)\eta_x]_x+[h_0(x,y)\eta_y]_y. 
\ea
Equation \eqref{926}, first derived by \cite{Tuck1972}, has analytical closed-form solution in one- and two-dimensional setups \cite[][]{Liu2003,Sammarco2008,Didenkulova2013}.

 \subsection{Wave energy in Lagrangian framework}
To derive the expression for wave energy in terms of Lagrangian variables we first consider the Eulerian definition of wave energy (normalized by $\rho g H^2 L^2$) in the domain $\Omega=\{(x,y,z)\in\mathcal{F}(t)\times[-h(x,y,t),\eta(x,y,t)]\}$ where $\mathcal{F}(t)\subset\mathbb{R}^2$ is the horizontal projection of the free-surface (that extends from one side's runup or rundown to the other side's runup or rundown):
\ba{B100}
\mathcal{E} = \mathcal{E}_{pot}+\mathcal{E}_{kin} = \int_{\mathcal{F}(t)}\int_{0}^{\eta} z \, \d z\, \d x \d y + \int_{\mathcal{F}(t)}\int_{-h}^{\eta} \f{1}{2} \lp u^2+v^2+\mu^2 w^2 \rp \d z\, \d x\d y.
\ea
Since $\mu=\frac{H}{L}\ll 1$, we see from \eqref{B100} that the  contribution of the vertical velocity is negligible. The potential energy $\mathcal{E}_{pot}$ readily reduces to
\ba{B101}
 \int_{\mathcal{F}(t) }\f{1}{2} \eta^2 ~ \d x\d y. 
\ea
Changing the variables of integration $(x,y,z)$ to the Lagrangian coordinates $(a,b,c)$ and noting that $\eta(x,y,t)=Z(a,b,0,t)$, $\vec{X}_t=(u,v,w)$, we rewrite the total wave energy as
\ba{B102}
\mathcal{E} = \f{1}{2} \int_{\mathcal F(0)} \lb \frac{\partial(x,y)}{\partial(a,b)}  Z^2(a,b,0,t) + \int_{-h_0(a,b)}^{0} \frac{\partial(x,y,z)}{\partial(a,b,c)}\lp X_t^2+Y_t^2+Z_t^2 \rp \d c \rb \d a \d b, 
\ea
where clearly the determinant $\frac{\partial(x,y,z)}{\partial(a,b,c)}=1$ because of continuity. Note that $\mathcal F(0)$ is the projection of the initial (flat) water surface and therefore gives the limits of integration for the Lagrangian variables $a,b$ in \eqref{B102}. Also it is to be noted that the vertical limit of integration is from $-h_0(a,b)$ to $0$ again because the integration is performed over Lagrangian variables. The final form of total wave energy is obtained once  Airy's expansion is substituted for the displacement variables. To the leading order, we get
\ba{B104}
\mathcal{E} = \f{1}{2} \int_{\mathcal F(0)} \lb \frac{\partial(x,y)}{\partial(a,b)} (Z^{(0)})^2 + h_0 \lp (X_{t}^{(0)})^2+(Y_{t}^{(0)})^2\rp \rb \d a\d b.  
\ea

%
\section{Numerical Implementation}
\subsection{Finite-difference scheme}
We solve the system of equations \eqref{927} by the finite difference method. Equation \eqref{927} is integrated in time using an explicit fourth-order method (Runge-Kutta 4) with $Z^{(0)}$ obtained from \eqref{928} and \eqref{929}. A compact finite-difference Pade scheme is implemented to obtain the first- and second-order spatial derivatives such that the accuracy of the numerical scheme is that of a $4^{th}$ order method. The computation of the derivatives on the edges of the domain depends on the type of the boundary. For a vertical wall, which acts as a mirror, the normal displacement is zero, and the tangential displacement has a vanishing normal derivative. 
For a shoreline, which we define as the intersection of a sloping beach with the free-surface, a forward or backward approximation is used in space to get the horizontal derivatives. The discretization of the derivatives on the boundaries is performed by adjusting the number of stencil points so that the order of approximation is the same as in the interior of the domain. 

We shall assume perfect reflection at the shore and require the solution to be bounded everywhere including near the moving shoreline. Since for the linearized problem on a plane beach one of the homogeneous solutions is proportional to the Weber function $Y_0$ in 2D (Bessel functions of the second kind) and to the Whittaker function $W_{n^2/2m,0}$ in 3D ($n,m \in \mathbb{N}$), which are both unbounded at the origin \cite[see e.g.][]{Shuto1967,Shuto1968}, small errors in numerical approximations can induce instability. Indeed for the 2D problem one obtains Bessel's equation for a plane sloping beach, i.e. by substituting $h(a)=a\tan \alpha$ in \eqref{919}, which presents a singularity at the shoreline $a=0$. To resolve this issue numerically, we integrate the governing equations up to one grid cell before the shoreline. As a result, the physical shoreline  is not part of the numerical domain, and is determined by a linear extrapolation  from the grid's boundary. The grid size is chosen small enough such that convergence is secured. 
\subsection{Model validation }\label{333}
{Numerical predictions of wave runup on a sloping beach based on our model equations are compared with analytical solutions of the celebrated hodograph-transformed shallow-water equations of \cite{Carrier1958}. Specifically, we consider an initial Gaussian waveform with zero velocity, i.e. case \textbf{a} in \cite{Carrier2003}, which in the Eulerian framework reads
\ba{301}
\eta(x,0) = H_{1} \exp(-c_{1}(x-x_{1})^{2}), ~~ \text{with} ~~ \eta_t(x,0) = 0.
\ea 
{Since case \textbf{a} of \cite{Carrier2003} (i.e. with $H_{1}=0.017$, $ c_{1} = 4.0,~  x_{1} = 1.69$) shows large steepnesses at the shoreline tip, which is not permitted under Airy assumptions, here we first compare our numerical predictions with the analytical solution obtained for an initial Gaussian waveform with a reduced amplitude of $H_{1}={0.017}/{2}$ but with $ c_{1},  x_{1}$ as before.}
Extreme values such as maximum runup and rundown are compared in table \ref{Carrier} for this case, referred to as \textbf{a'}, and a very good agreement is obtained.}
\begin{table}
\centering
	\begin{tabular}{ c | c | c | c | c  } 
Case & 	Max. runup & Min. rundown & Max. shoreward speed & Max. seaward speed \\ \hline
\textbf{a'} \begin{tabular}{c} (i) \\ (ii) \end{tabular} & \begin{tabular}{c} 0.02343 \\ (0.0235) \end{tabular} & \begin{tabular}{c} -0.01352 \\ (-0.0134) \end{tabular} & \begin{tabular}{c} -0.05043 at $x$=-0.0142 \\ (-0.0514 at $x$=-0.0142) \end{tabular} & \begin{tabular}{c} 0.1087 at $x$=0.000856 \\ (0.1066 at $x$=0.000376) \end{tabular} \\ \hline
\textbf{a} \begin{tabular}{c} (i) \\ (ii) \end{tabular} & \begin{tabular}{c} 0.0467 \\ (0.0470) \end{tabular} & \begin{tabular}{c}  -0.0272 \\ (-0.0268) \end{tabular} & \begin{tabular}{c}-0.0991 at $x$=-0.0258 \\ (-0.103 at $x$=-0.0260) \end{tabular} & \begin{tabular}{c} 0.222 at $x$=0.0153 \\ (0.213 at $x$=0.0122) \end{tabular} \\ \hline
\end{tabular}      
	\caption[]{Comparison of wave runup extrema on an idealized beach ($h(x)=x$) between: (i) numerical simulations of equation \eqref{919} and (ii) the analytical solution derived by \cite{Carrier2003} (in parentheses). The initial Gaussian waveform is described by \eqref{301} with amplitude $H_{1} = 0.017/2$ for case \textbf{a'} and $H_{1} = 0.017$ for case \textbf{a}. Other parameters are $ c_{1} = 4.0,  x_{1} = 1.69$. The analytical results for case \textbf{a} are the ones reported in \cite{Carrier2003} (with sign of the runup/rundown changed to correspond to runup heights instead of penetration depths), while those of case \textbf{a'} are based on a numerical integration of equation (2.9) in \cite{Kanoglu2004}. The numerical simulations results are given for $\delta a = 2\times 10^{-3}$, $\delta t = 4.88 \times 10^{-5}$.}	
	\label{Carrier}
\end{table}
{Although large wave steepnesses at the shoreline invalidates our model assumptions \cite[see the discussion of][on this]{Meyer1986,Meyer1986a}, we still compare our numerical predictions with the analytical solution for the original case \textbf{a} of \cite{Carrier2003}. The Jacobian of the hodograph transformation has been shown to change sign very close to the shoreline in this case \cite[cf. figure 17 in][]{Carrier2003}, which is an indicator of wave breaking. It was, however, remarked by \cite{Synolakis1987}, who conducted laboratory experiments on solitary waves, that the post-breaking analytical wave profile approximates the actual wave very well when wave breaking occurs very close to the shoreline. The fact that we obtain a very good match in table \ref{Carrier} for the original case \textbf{a} of \cite{Carrier2003} suggests that numerical simulations may continue to predict accurate wave shape despite having waves breaking at the shoreline. This is of course permitted by the fact that we moved the first grid point  away from the line of vanishing depth.} {We would like to point out that in our direct simulation of the governing equation \eqref{919} for the case \textbf{a} the wave steepness which is given by 
\ba{near break} 
{\eta}_{{x}} = \lb\f{z_a(a,c,t)}{x_a(a,c,t)}\rb_{c=0} = \lb\f{Z_a(a,c,t)}{1+X_a(a,c,t)}\rb_{c=0}
\ea
grows unbounded as we decrease the grid size (figure \ref{Conv}a), i.e. wave steepness at the shoreline does not converge. This is not unexpected since the Jacobian of the hodograph transformation in this case vanishes at the shoreline as discussed before. The runup itself however converges to the true runup value. For case \textbf{a'}, i.e. with decreased wave amplitude, the Jacobian does not vanish. Correspondingly, in our simulations, both the maximum runup value and wave steepness converge asymptotically (c.f. figure \ref{Conv}b).} 
 
 In the following section (\S4) we study the evolution of landslide tsunamis in a lake, and present examples that highlight the effect of nonlinearities and wave reflections as well as interactions in the inundation maps that are specific to lakes and enclosed bodies of water.

\begin{figure}
\centering
\includegraphics[height=4.4cm]{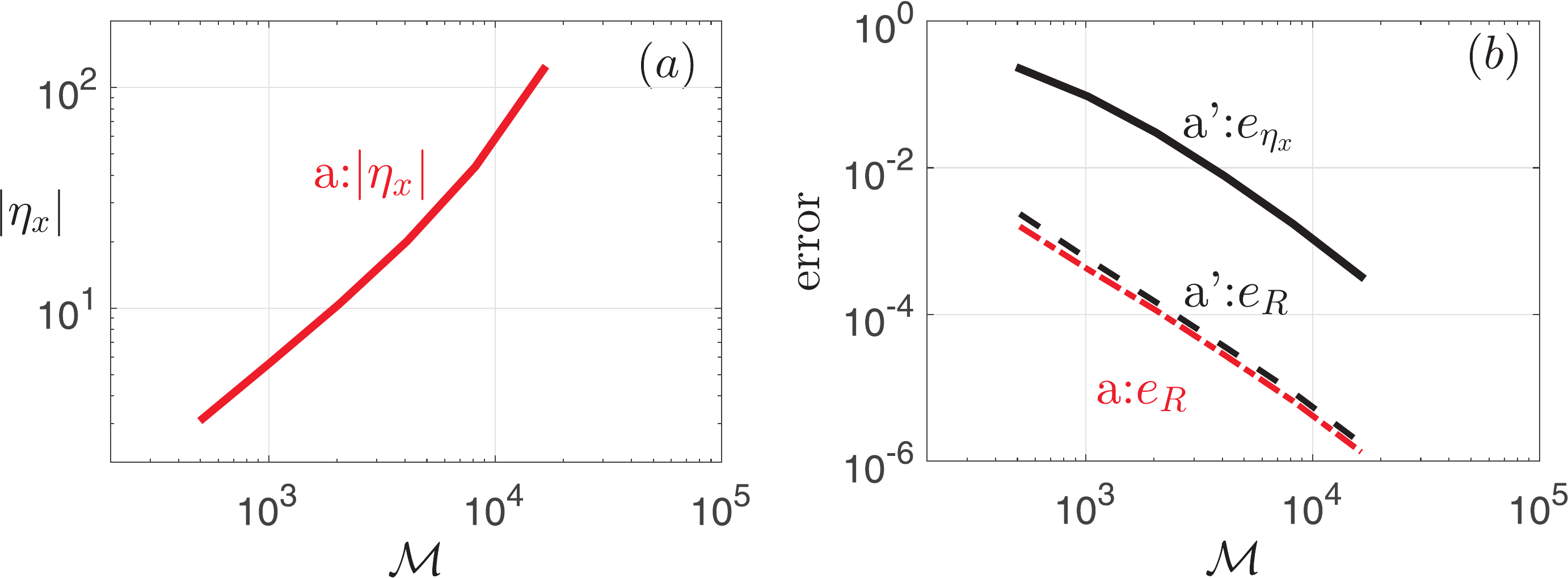} 
\caption{(a) The absolute value of the maximum wave steepness at the shoreline (\red\L) as a function of $\mathcal{M}$ for case \textbf{a}. Clearly in this case $|\eta_x|$ grows exponentially with $\mathcal{M}$ and does not converge. (b) The relative error of the maximum wave runup $e_R$ (\L), and the relative error of the maximum wave steepness $e_{\eta_x}$ (\dashL) at the shoreline are shown as functions of the number of grid points $\mathcal{M}$ for the numerical simulation of case \textbf{a'} reported in table \ref{Carrier}. These relative errors are defined with respect to the values obtained for $\mathcal{M}=2^{15}$ (i.e. the finest grid simulated). The relative error of the maximum wave runup for case \textbf{a} (\red\dashdot) is also shown in this plot. }
\label{Conv}
\end{figure}
\section{Landslide Tsunamis in Lakes}
\subsection{The Setup}

We focus our attention on a geometrically simple lake to highlight some of the physics involved. Specifically, we consider a shallow lake of rectangular surface area $L_a\times 2L_b$ confined symmetrically by two opposing vertical walls along $y=\pm L_b$ and two opposing sloping beaches aligned with the $\pm x$ direction (figure \ref{RefSpace}). A vertical wall  may be an idealization of a  mountain cliff or a dam. We assume that the two opposing beaches have the same slopes, and that they each occupy 1/3 of the horizontal extent of the lake. The remaining middle 1/3 is flat with the dimensionless depth of unity therefore the beach slope is $\tan\alpha=3/L_a$ in dimensionless space and $\mu \tan\alpha$ in physical domain (note that in the dimensionless space vertical and horizontal lengths are scaled differently, c.f. \eqref{852}). The bottom is therefore given by
\ba{930}
h_0(x,y) = \left\{
\begin{array}{ccc}
x\tan \alpha, & 0<x<L_a/3,~-L_b<y<L_b, \\
1, & L_a/3<x<2L_a/3,~-L_b<y<L_b, \\
3 - x\tan \alpha, &  2L_a/3<x<L_a,~-L_b<y<L_b.
\end{array} \right. 
\ea
To avoid sharp corners an arc of radius $1/20$ (in dimensionless variables) connects each sloping beach to the flat bottom.

\begin{figure}
\begin{center}
\includegraphics[width=0.6\textwidth]{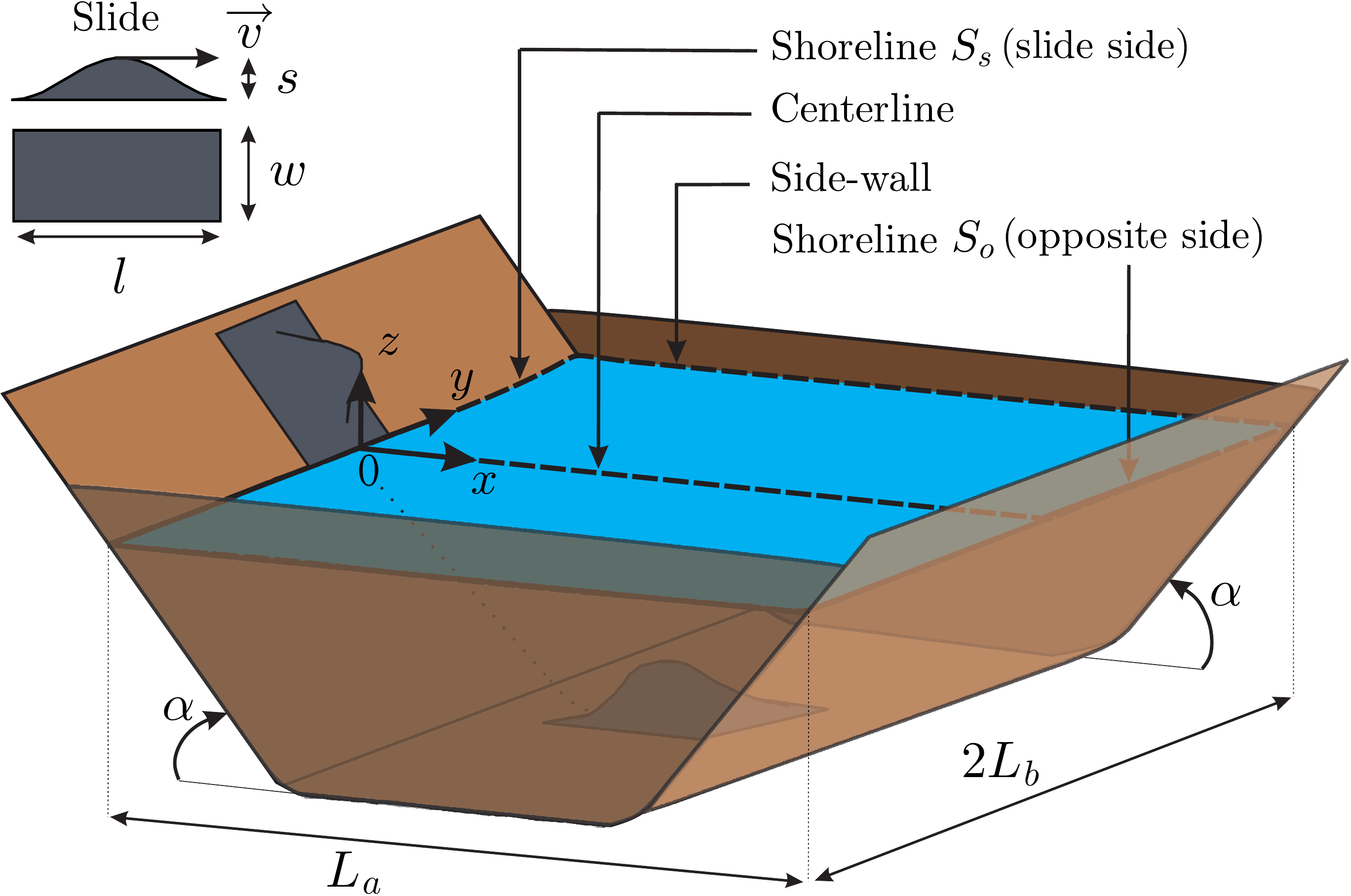}
\end{center}
\caption{Schematic of the lake and the landslide. The landslide is shown at both its initial position (above water) and its final position (underwater). }\label{RefSpace}
\end{figure}

 We consider the motion of a landslide whose height is given by 
\ba{931}
h_{s}(x,y,t) = \left\{ \begin{array}{c}
- s\sin^2\lb \f{\pi}{l} (x-v t)\rb\cos^2\lp\f{\pi}{w}y\rp,~\textrm{for}~~-l<x-vt<0,~~-\f{\omega}{2}<y<\f{\omega}{2},\\
0, \text{ otherwise, } \end{array} \right.
\ea
which describes a smooth and rigid hump-like landslide with a horizontally projected surface area of length $l$, width $w$ and  maximum thickness $s$ that moves along the lake's centerline (figure \ref{RefSpace}). The total water-depth is therefore $h(x,y,t)=h_0(x,y)+h_{s}(x,y,t)$. 
 Physically speaking, \eqref{931} corresponds to a slide moving with a constant speed $v\sqrt{1+\tan^2 \alpha}$ down the beach. When the center of the slide reaches the beach's toe at $x=L_a/3$, it decelerates according to a smooth cubic law $v \propto t^3$ until it reaches the middle of the lake where it stops (i.e. at $x=L_a/2$, c.f. figure \ref{RefSpace}). 

Two types of waves are generated  by the 3D subaerial landslide as it enters the water. We refer to waves that propagate freely across the lake as {\em outgoing waves}, and to waves that are trapped by the shoreline as {\em edge-waves }.  Outgoing waves along open coasts, forced impulsively  by landslides, have been the subject of a large number of laboratory experiments conducted for both solid landslides \cite[e.g.][]{Kamphuis1970,Walder2003,Panizzo2005,Panizzo2005a,Heller2012} and granular landslides \cite[][]{Fritz2004,Ataie-Ashtiani2008,DiRisio2011}. More recently, properties of landslide-generated edge-waves have also been investigated \cite[e.g.][]{Lynett2005,Liu2005,Sammarco2008,DiRisio2009}. The following analysis of landslide tsunamis in lakes highlights the importance of nonlinearity as well as the significance of the interactions between outgoing and edge waves in the generation of unexpectedly large runups.

\subsection{Numerical Results and Discussion}

We choose the lake dimensions $L_a=1.00$, $L_b=1.02$ and normalized slopes of $\tan \alpha = 3$. A solid mass of $w = l = 0.25$ with thickness $s=0.07$ according to \eqref{931} slides down the beach $S_s$ with a horizontal speed $v=0.12$ (this specific set of parameters is chosen as it further highlights the physics to be discussed. A detailed sensitivity analysis is provided in \S4.3). Simulation parameters are chosen to be $\delta a = 0.002$, $\delta b = 0.002$, with Courant number equal to 0.5 such that $\delta t = 0.001$, for which our simulations converge.

\begin{figure}
\hspace{-0.4in}
\includegraphics[width=1.25\textwidth]{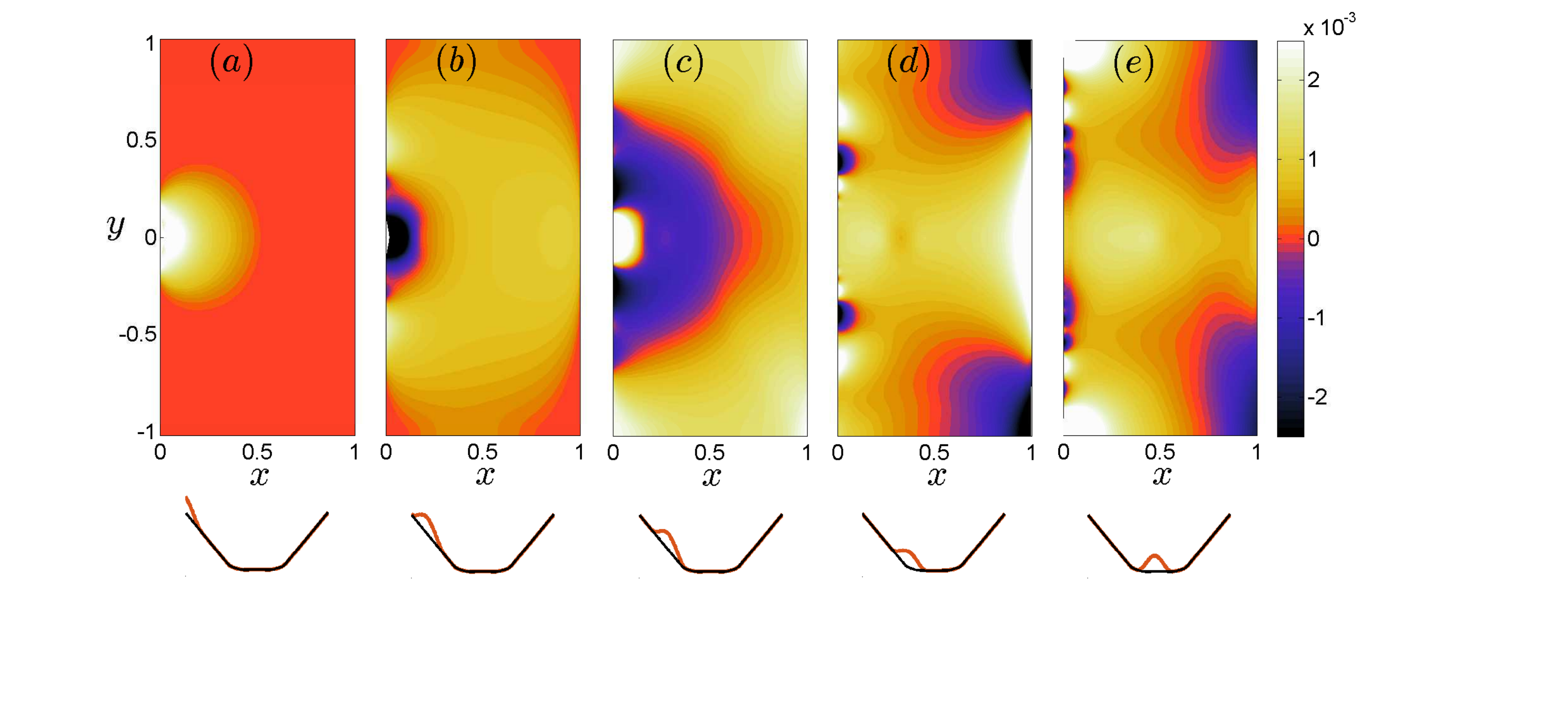} \vspace{-0.55in} 
\caption[]{Snapshots of nonlinear dimensionless free-surface height $Z$ at ($a$) $t=0.83$, ($b$) $t=1.66$, ($c$) $t=2.47$, ($d$) $t=3.76$, ($e$) $t=5.63$.  Physical parameters associated with the lake's dimensions and slide's shape are $s$=0.07, $v$=0.12, $L_{b}=1.02$, $w = l = 0.25$, $\tan \alpha = 3$. Simulation parameters are $\delta a = 0.002$, $\delta b = 0.002$, $\delta t = 0.001$. The position of the slide with respect to the lake geometry is shown below each plot (slide height is magnified by a factor of four).} \label{Snapshots}
\end{figure} 
 
{Snapshots of the water surface are shown in figures \ref{Snapshots}a-e. Lighter colors show higher elevations and darker colors show depressions. Upon water entry at $t=0$ the land mass pushes the water forward, generating a big hump of water at its front (figure \ref{Snapshots}a). This first hump then disintegrates into a radially-spreading outgoing wave and a series of shoreline-trapped edge-waves on the slide-side $S_s$. A pair of edge-wave crests can be seen in figure \ref{Snapshots}b traveling away from the centerline $y$=0. The outgoing wave crosses the length of the lake and reaches the opposite shoreline $S_o$ (figure \ref{Snapshots}b), while a ring of depression is formed near the origin on the $S_s$ shore. Moments later (figure \ref{Snapshots}c), the first outgoing- and edge-waves reach the corners of the lake ($y=\pm L_b$) while a second  hump is formed at the origin. Figure \ref{Snapshots}d shows the runup of the second outgoing-wave on $S_o$, and the propagation of the edge-waves (resulting from the disintegration of the rebound hump) along $S_s$ . The leading edge-wave of this second hump is the longest and largest in amplitude, as  seen in white at $y \sim 0.63$ in figure \ref{Snapshots}d. Finally, figure \ref{Snapshots}e shows the return of the second outgoing wave (the one initiated on $S_s$ in figure \ref{Snapshots}c)  to $S_s$ after reflection from $S_o$. This reflected wave superimposes on the trail of edge-waves on $S_s$,  resulting in higher runups at some locations along $S_s$, as will be elaborated shortly.}

The time history of the water surface evolution along   the shoreline $S_s$ ($x$=0) is shown  in figure   \ref{TimeHis}a,   the shoreline $S_o$ ($x$=$L_a$) in \ref{TimeHis}b,    the centerline of the lake ($y=$0) in \ref{TimeHis}c, and the vertical side-walls ($y=\pm L_b$) in \ref{TimeHis}d. The first runup due to the water entry of the landslide is seen in figure \ref{TimeHis}a at $y=0, t\sim 0.5$. It is then followed by a rundown that starts at $t \sim 1.4$. From figure \ref{TimeHis}a it is seen that the dispersion of edge-waves along $S_s$ is similar to that resulting from an initial hump in open water: longer waves travel faster and more crests are generated as time goes by \cite[c.f.][]{Sammarco2008}. Due to dispersion, edge-waves elongate and accelerate as they travel away from the origin \cite[c.f.][]{Sammarco2008,Mei2005}, as can be seen from the convex shape of the edge-wave pathlines in figure \ref{TimeHis}a (marked by solid and dashed lines).

{Figure \ref{TimeHis}a also shows the reflection of edge-waves by the vertical side-walls (i.e. $y=\pm L_b$). The second edge-wave due to the first runup (solid line) meets the leading edge-wave due to the second runup (dashed line) when reaching the side-wall at $y=L_b,t\sim 4.1$. It then bounces back to interact with the other forward moving (i.e. $+y$ direction) edge-waves of the second hump (marked by dash-dotted lines). A relatively high runup at $y\sim 0.61, t\sim 4.8$ (highlighted by a green circle) is partially due to this encounter.}

{The radially outgoing wave first reaches the opposite shoreline $S_o$ at $y=0,t \sim 1.7$ (figure \ref{TimeHis}b). Wave runup patterns along $S_o$ (solid white and black lines) are arc-shaped due to radial spreading. The leading crest (marked by the white line) is  reflected from the side wall $y=L_b$ near $S_o$ at $t\sim$ 2.7 and turns into an edge-wave whose path is marked by a black dashed line. This edge-wave advances toward the center, and interacts with the second runup (marked by a black solid line) at $t\sim $ 3.7, resulting in a constructive interference (as marked by a circle). These interactions continue along $S_o$ as more edge-/outgoing-waves come into play.} 
\begin{figure}
\hspace{-0.15in}
\includegraphics[width = 1.1\textwidth]{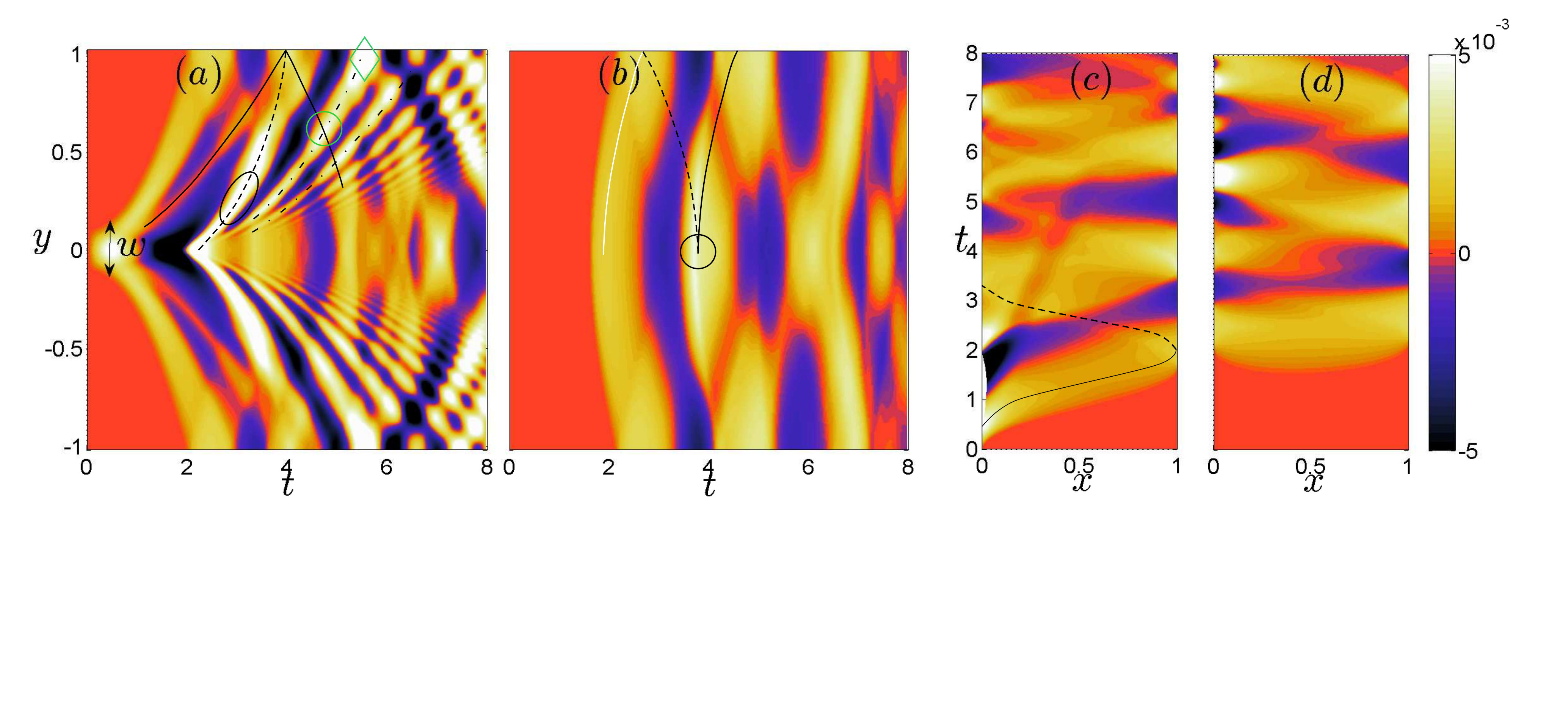} \vspace{-0.95in}
\caption[]{Time history of the nonlinear wave runup on (a) shoreline $S_s$, (b) shoreline $S_o$, and wave height along (c) centerline, (d) side-wall, for a landslide tsunami with physical and simulation parameters given in figure \ref{Snapshots}. The color map shows the dimensionless free-surface height $Z$. }\label{TimeHis}
\end{figure}
Contrary to edge-waves, outgoing waves are nondispersive. Therefore, as seen in figure \ref{TimeHis}c, the depression and elevation rays appear straight where the water depth is constant (i.e. $1/3 \leq x \leq 2/3$), and curved near the shorelines because of refraction. The superposition of the leading edge-wave of the rebound (second) runup on $S_s$, with the reflected first outgoing wave (marked by a dashed line in figure \ref{TimeHis}c), can be clearly seen in  figure \ref{TimeHis}a. The bright white spot resulting from this interaction is highlighted by an ellipse at $t\sim 3.3$. Figure \ref{TimeHis}c shows that the two outgoing crests (from the first and the second runups) remain separated for the rest of the simulations, and that the amplitude of the first outgoing-wave is smaller than the second one. Figure \ref{TimeHis}d shows the surface elevation along the vertical side walls. Both edge-waves and outgoing waves emanating from the first runup reach the side-walls at approximately $t\sim 2.7$. They generate an almost horizontal beam across each of the two vertical walls. Edge-waves on $S_s$ ($x=0$) are also seen  in figure \ref{TimeHis}d with their crests at $t \sim 4, 5.5,7.5$,  and on $S_o$ ($x=L_a$) at $t \sim 4.5,6.5$.

The multiple interactions between reflected outgoing-waves and edge-waves that occur in a lake suggest that extreme inundations may take place far away from the immediate neighborhood of the landslide and some time after its submergence. As an example, in our simulation, an extremely large runup occurs at the lake's corners ($x=0,y=\pm L_b$) due to constructive interference between the third edge-wave of the initial runup, the second edge-wave of the rebound (second) runup, and the reflected second outgoing-wave. This runup takes place at $t=5.5$ and is almost twice as big as any other runup along $S_s$ (seen as a bright-colored region in figure \ref{TimeHis}a marked by a diamond). This runup is also much larger than that estimated by an open-coast calculation, as can be seen in figure \ref{open}a where we show the wave runup at $x=0,y=1.02$ as a function of time due to the same slide entering four different geometric configurations:  (1)- a lake of finite size ($L_a=1,~ L_b=1.02$, solid red line), (2)- an infinitely long lake ($L_a\rightarrow \infty$,  blue dashed line), (3)- an infinitely wide lake ($L_b\rightarrow \infty$, beige dotted line), and (4)- an open coast ($L_a,L_b\rightarrow \infty$, green dash-dotted  line). As expected, the runup is greatest for the finite-sized lake due to the presence of a wall at $y=L_b=1.02$ and an opposite beach at $x=L_a=1$. The maximum runup for the case of an infinitely long lake, infinitely wide lake and an open coast is respectively 48\%, 54\% and 71\% lower than that of a finite size lake considered here.
\begin{figure}
\centering
\includegraphics[height=0.29\textwidth]{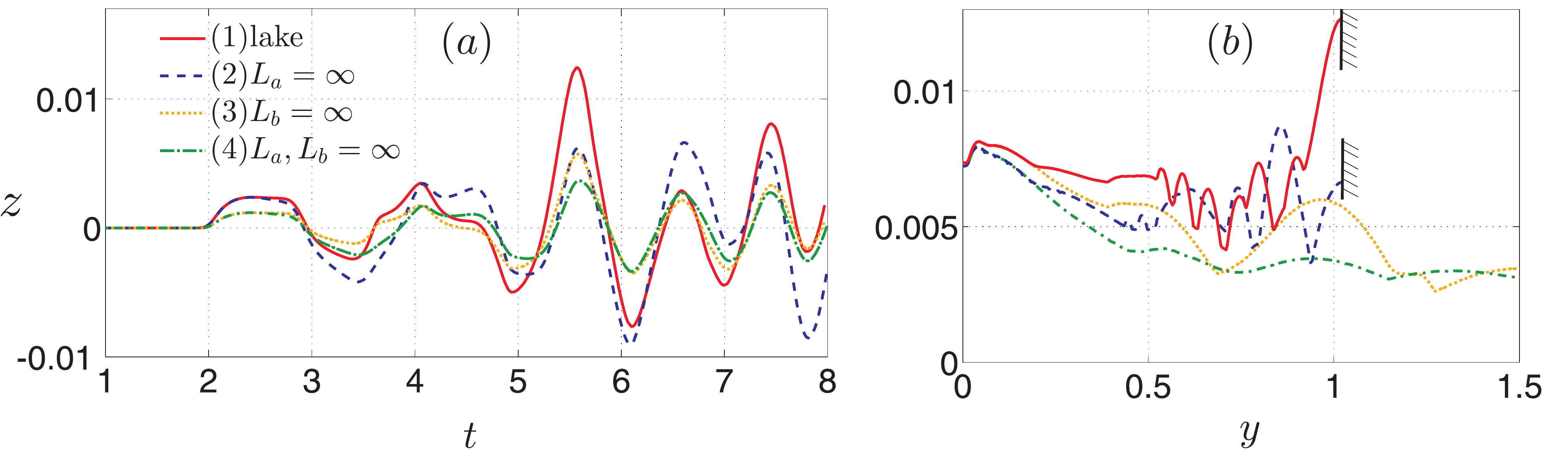} 
\caption[]{(a) Time history of nonlinear wave runup height at $y=1.02$ on $S_s$ for a slide entering: (1) a finite rectangular lake ($L_a=1$, $L_b=1.02$, \red\L), (2) a long and narrow lake ($L_a=\infty$, $L_b=1.02$, \blue\dashL), (3) a short and wide lake ($L_a=1$, $L_b=\infty$, {\color{orange}\dotL}), (4) an open coast ($L_a=\infty$, $L_b=\infty$, {\color{green}\dashdot}). (b) Inundation maps up to $t=8$ along $S_s$ for $y\geq0$ for  the four cases displayed in figure (a). Physical and simulation parameters other than $L_a$ and $L_b$ are the same as in figure \ref{Snapshots}.}
\label{open}
\end{figure} 

The inundation map of shore $S_s$ ($y\geq0$) for the four geometries considered above shows (figure \ref{open}b) that  the average runup increases in the presence of a wall and an opposite beach . For the open coast (green dash-dotted line), edge-waves result in a large runup very close to the origin less than a quarter of the slide's width away from the centerline. Edge-waves amplitude then decreases as they propagate farther away \cite[which is in qualitative agreements with earlier studies by][]{Lynett2005,Sammarco2008,DiRisio2009}. The contrast between the four cases studied here shows the importance of the wave reflection from side/opposite boundaries and the resulting superposition in the the magnitude and the location of the maximum runup.

Of practical  interest is also the importance of nonlinearities in the simulation presented. The inundation maps of the two shorelines $S_s$ and $S_o$, as well as the maps of maximum wave height along the lake centerline and side-walls, can be used to measure the significance of nonlinearity in the runup predictions. Figure \ref{nonlinab}a compares the linear vs nonlinear runup on both $S_s$ and $S_o$.  The maximum inundation predicted by the nonlinear theory is higher than the linear prediction everywhere on both $S_s$ and $S_o$. In particular, the relative increase due to nonlinear effects at $y \sim 0.05$ on $S_s$ is $ \sim 50\%$. Details of inundation curves can be better understood in view of figures \ref{TimeHis}. For instance, the six local maxima between $y=0.55$ and $y=0.9$ on $S_s$ shown in  figure \ref{nonlinab}a are partially due to the reflections by the side-wall at $y=L_b$ of the third and fourth edge-waves of the rebound runup interacting with trailing edge-waves at $t \sim 7$. The extreme corner runup is found 20\% higher with nonlinear effects taken into account.

{On the opposite shore $S_o$, the runup of the second outgoing wave  is responsible for most of the inundation for $y < 0.5$ and is about 30\% higher when nonlinear terms are taken into account. The effects of wave shoaling and radial spreading are clearly seen in figure \ref{nonlinab}b by the rapid decrease of maximum wave height along the centerline away from the shorelines.} {Clearly, the quantitative difference between linear and nonlinear inundations is maximum closer to the shoreline since, due to nonlinear effects, larger wave height and steepnesses both contribute to a larger runup \cite[c.f.][]{Didenkulova2007b}}.

\begin{figure}
\centering
\includegraphics[height=0.35\textwidth]{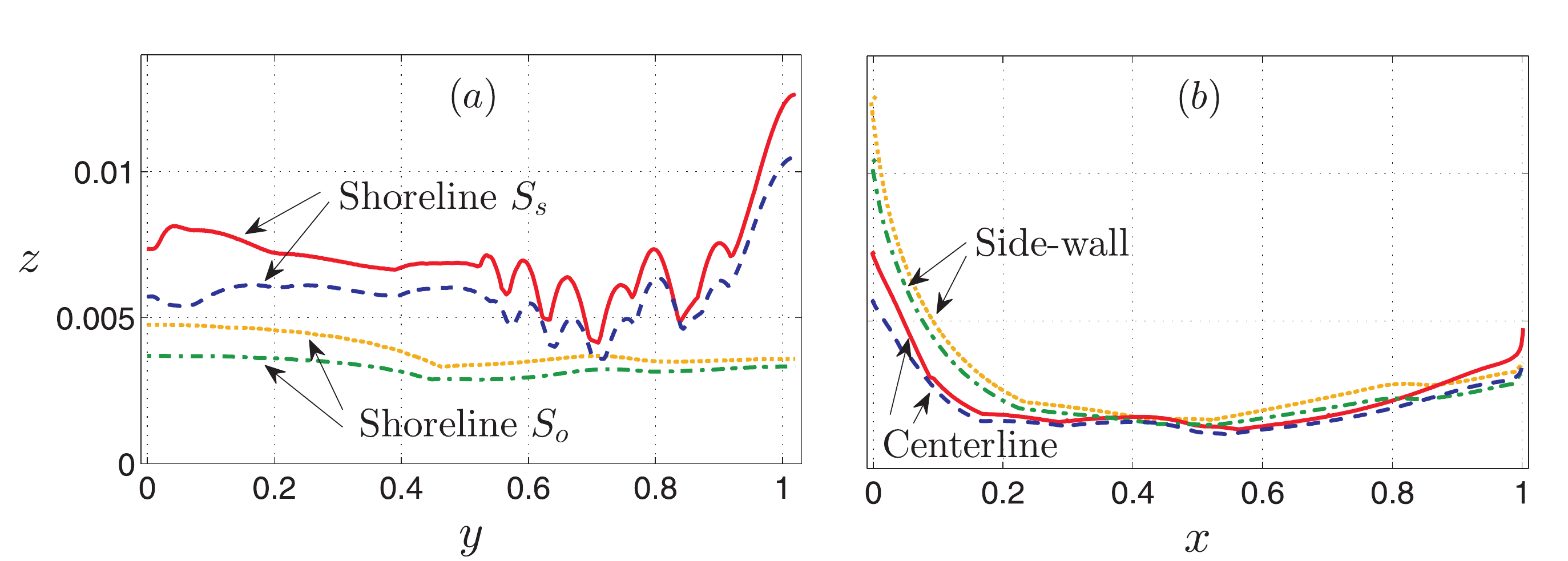}
\caption[]{Inundation maps of the (a) shorelines $S_s$, $S_o$ (for $y \geq 0$), and, (b) side-walls and centerline as predicted by nonlinear (\red\L ; {\color{orange}\dotL}) and linear theories (\blue\dashL ; {\color{green}\dashdot}). Physical and simulation parameters are the same as in figure \ref{Snapshots}. Simulation are performed from $t$=0 to $t$=8 beyond which no further significant changes occur.}
\label{nonlinab}
\end{figure} \noindent

To investigate at what time stage during the landslide event major nonlinear effects come into play and where they are highlighted, we compare the maximum linear and nonlinear wave heights for all surface fluid particles (i.e. all pairs $(a,b)$) at five different times (figure \ref{Nonlin}a-e). To quantify this difference between nonlinear and linear predictions we define 
\be \label{51}
\mathcal{N}(a,b,t) = \f{Z_{NL}(a,b,t) - Z_{L}(a,b,t)}{Z_{L,max}}\times 100 \ee
where $Z_{NL}$ ($Z_L$) is the maximum nonlinear (linear) wave-height for each pair ($a,b$) and time $t$, and $Z_{L,max}=\max\lb Z_{L}(t=t_f\rightarrow \infty) \rb$ is a single number that shows the overall maximum wave height as predicted by the linear theory. In our simulations the final time $t_f$=8 is chosen as $Z_{L,max}$ no longer changes for $t>t_f$. Stronger differences (in \%) are shown with darker colors. Note that the linear results are based on equation \eqref{925}.

Nonlinearity comes, as is expected, from wave-bottom interactions near the line of vanishing depth (i.e. near the shoreline). Yet, the first runup on $S_s$, which is due to the slide starting to push the water, is only weakly nonlinear (as seen in figure \ref{Nonlin}a). Therefore, major nonlinearities come into play when the rebound hump is formed and disintegrated into edge-waves and outgoing-waves (see dark patch localized near $S_s$ in figure \ref{Nonlin}b). Physically speaking, this is due to the rundown being amplified by a downward dragging imposed by the landslide descending further under the water. Excess of waveheight predicted by nonlinear terms are then seen almost everywhere in the lake as edge waves and outgoing waves reach the opposite shoreline and sidewalls (figure \ref{Nonlin}c,d). Finally, figure \ref{Nonlin}e shows the contrast between maximums of nonlinear and linear predictions from $t=0$ to the final time $t_f=8$. As discussed before, nonlinearity is much stonger near the shorelines and at the corners of the lake.

We remark that the strong contrast between nonlinear and linear predictions is partly due to the three-dimensional nature of the landslide-generated waves. To demonstrate this, we compare in figure \ref{NonDim} nonlinear and linear maximum waveheights along the lake's centerline (i. e. $\mathcal{N}(a,0,t)$) for the finite-width slide (i.e. $w < L_b$; beige solid line) of the case studied in figures \ref{Snapshots}-\ref{Nonlin}, and an infinitely-wide landslide (ie. $w \rightarrow \infty$; blue dashed line), which makes the waves two dimensional. As discussed before the large nonlinearity $\N$ near $a$=0 for the 3D case in figure \ref{NonDim}b appears with the second runup ($\mathcal N(a,0,1.6)$=16\%). The much smaller difference in the 2D case ($\mathcal N(a,0,1.6)$=5\%) thus suggests that this increase is due to three-dimensional effects. Note that $\N$ is even larger than 16\% in the 3D case just slightly away from the centerline ($\mathcal N(a,0.05,1.6)$=28\%, c.f. figure \ref{nonlinab}). Another interesting behavior specific to the 2D case is that the linear theory overpredicts the maximum waveheight sometime after the slide submergence (figure \ref{NonDim}d,e near the opposite shoreline). Note that the absolute value of $Z_{L,max}$ in the 2D case can be much larger (about four times in the presented case) than its 3D counterpart which is due to energy spreading in the 3D case.

\begin{figure} 
\hspace{-0.15in} 
\includegraphics[width=1.25\textwidth]{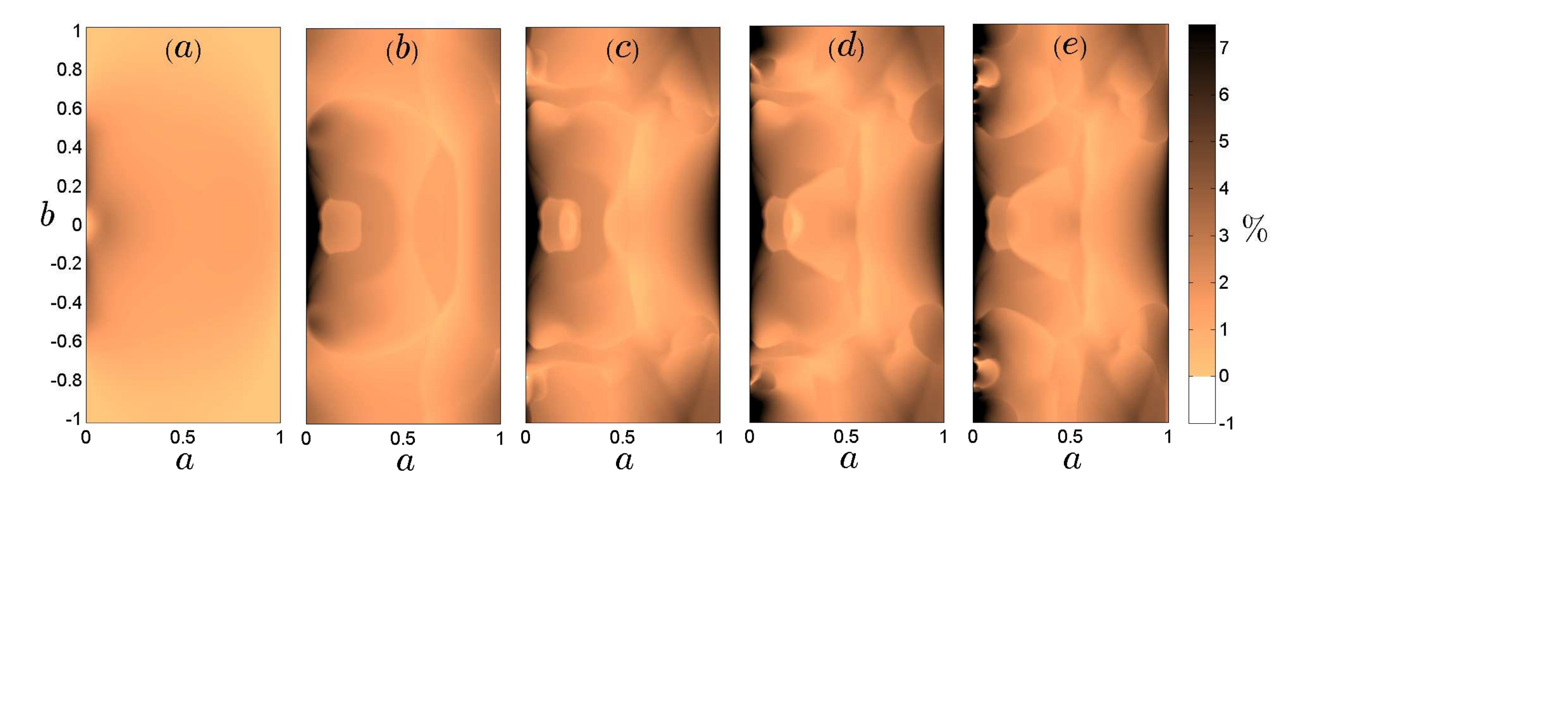}  \vspace{-1.25in} 
\caption[]{Spatial distribution of the difference between nonlinear and linear maximum waveheights ($\mathcal N$, c.f. \eqref{51}) during a landslide tsunami in a lake. Figures a-e show the difference during time periods lasting from zero to $t_f$=1.64(a), 3.30(b), 4.97(c), 6.63(d), 8.00(e). During the entire event, nonlinear waveheight predictions are higher than the linear ones, and the difference is higher close to the shorelines (c.f. figure \ref{nonlinab}). Physical and simulation parameters are the same as in figure \ref{Snapshots}.} 
\label{Nonlin}
\end{figure} 
\bigskip
\begin{figure}
\hspace{-0.3in}
\includegraphics[width = 1.2\textwidth]{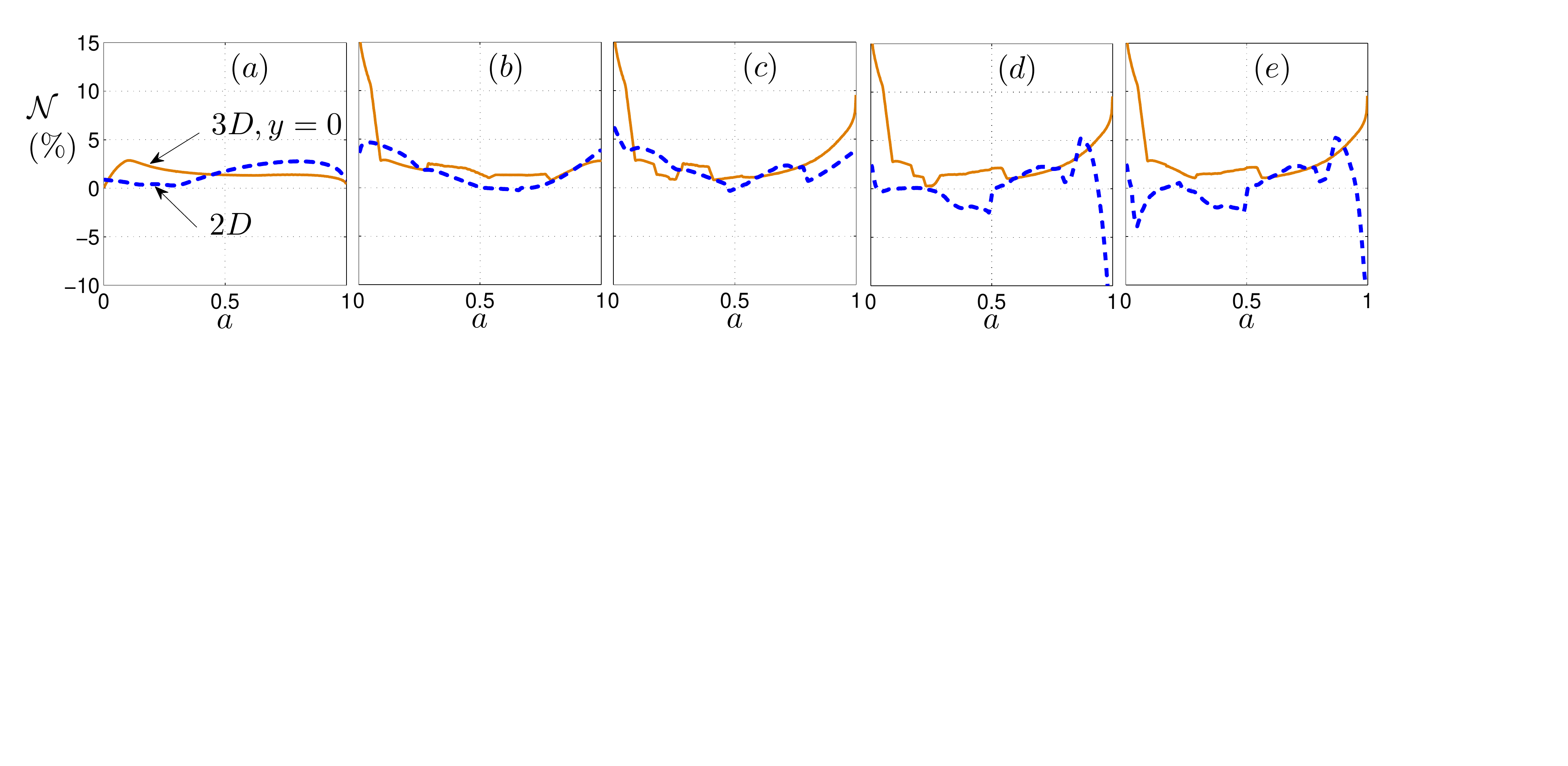}
\vspace{-2in}\caption[]{Effect of the slide's width on the difference between nonlinear and linear maximum waveheights along the lake's centerline. The predictions are plotted for a finite width landslide ($w<L_b$, {\color{orange}\L}) and an infinitely-wide landslide ($w\rightarrow \infty$,\blue\dashL). In the latter case the problem is two-dimensional, and therefore this figure also highlights effects of three dimensionality on the significance of nonlinearity.  Figures a-e show the difference ($\mathcal N$, c.f. \eqref{51}) for time periods from zero to $t_f$ = 1.64, 3.30, 4.97, 6.63, 8.00 (i.e. same as in figure \ref{Nonlin}). Note that $\mathcal N$ is much higher for the 3D case, and that $\mathcal N$ becomes negative at some time in the 2D case. Physical parameters (other than $w$) and simulation parameters are the same as in figure \ref{Snapshots}.}
\label{NonDim}
\end{figure}

It is of interest also to assess how the energy ceded by the slide to the fluid is divided between edge-waves and outgoing waves. For this purpose, in figure \ref{energy} we plot the total energy in the lake (black solid line), total energy very close to the shoreline $S_s$ in the area $0 \leq a \leq 1/10$ (dashed blue lines), and the total energy in the rest of the lake, i.e. in the area $1/10 \leq a \leq 1$ (dash-dotted red line). The total energy is calculated via \eqref{B104} and normalized by total wave energy in the steady state when $t\rightarrow \infty$ (here $t=8)$. The plot in figure \ref{energy}a is with the same parameters as in figures \ref{Snapshots}. For comparison, we also show the spatial energy division for a smaller slide $s=0.07/2$ that moves: (i) at the original speed $v=0.12$ (figure \ref{energy}b), and (ii) twice as fast (i.e. $v=0.24$, figure \ref{energy}c). In all three cases (the first two being qualitatively very much alike) there is  an overshoot of the total energy. This is because the initial outgoing-wave has time to reflect on $S_o$ back to $S_s$ while the slide is still moving forward and opposes the motion of the reflected wave. The total energy reaches its steady-state value when the slide stops moving, i.e. at $t=6$ for the two slow slide cases and $t=3$ for the faster slide.

Comparison of figures \ref{energy}a,b and \ref{energy}b,c show that the slide thickness does not have a major effect on the energy distribution while slide velocity does. For the slow landslide (figure \ref{energy}a,b) more energy is given to the outgoing waves and less energy is trapped, while for the fast slide the energy partitioning is almost even (figure \ref{energy}c). In figure \ref{energy}b, the first two humps of nearshore energy at $t\sim$0.9, 1.9 correspond to the generation of the first and second outgoing waves whereas the last three are associated with the return to $S_s$ of: (i) the first outgoing wave at $t=3.3$ and (ii) the second outgoing wave at $t=5.4$ and $7.7$. 

\begin{figure}
\hspace{-0.25in}\includegraphics[height=0.55\textwidth]{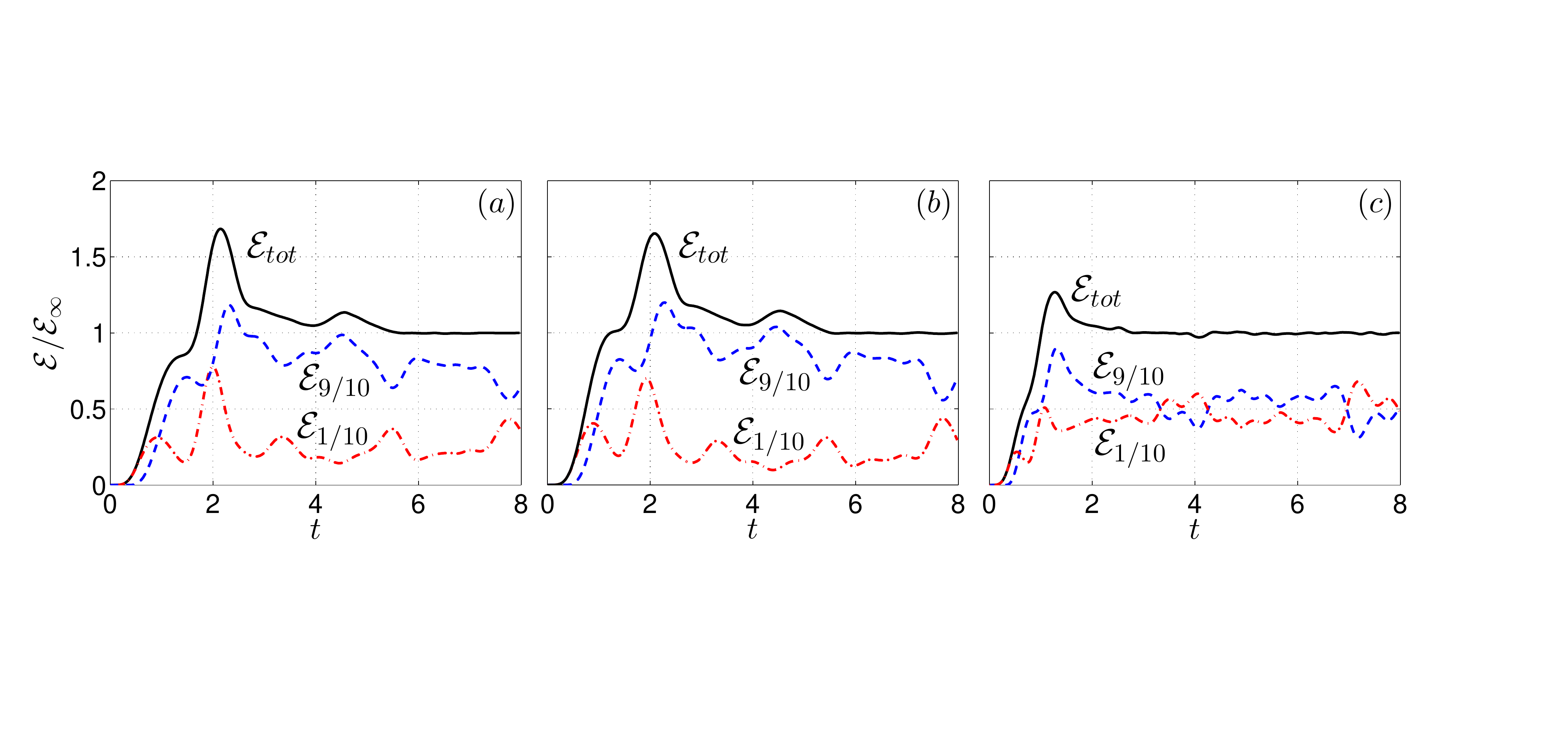}
\vspace{-1.in}\caption[]{Wave energy in the lake ($\mathcal{E}$, c.f. \eqref{B104}) normalized by steady-state wave energy ($\mathcal{E}_{\infty}$) is shown for a landslide with: ($a$) $s=0.07$, $v=0.12$ ($\mathcal{E}_{\infty}=0.967\times 10^{-6}$); ($b$) $s=0.035$, $v=0.12$ ($\mathcal{E}_{\infty}=0.229 \times10^{-6}$); ($c$) $s=0.035$, $v=0.24$ ($\mathcal{E}_{\infty}=2.220\times10^{-6}$). The total wave energy in the lake (\L) is divided to energy in the near $S_s$ area ($0<a<L_a/10$, \red\cdotL); and the rest of the lake ($L_a/10<a<L_a$, \blue\dashL). Other physical and simulation parameters are given in figure \ref{Snapshots}.  }
\label{energy}
\end{figure}

\begin{figure}
\begin{center}
\hspace{-0.in} 
\includegraphics[height=0.27\textwidth]{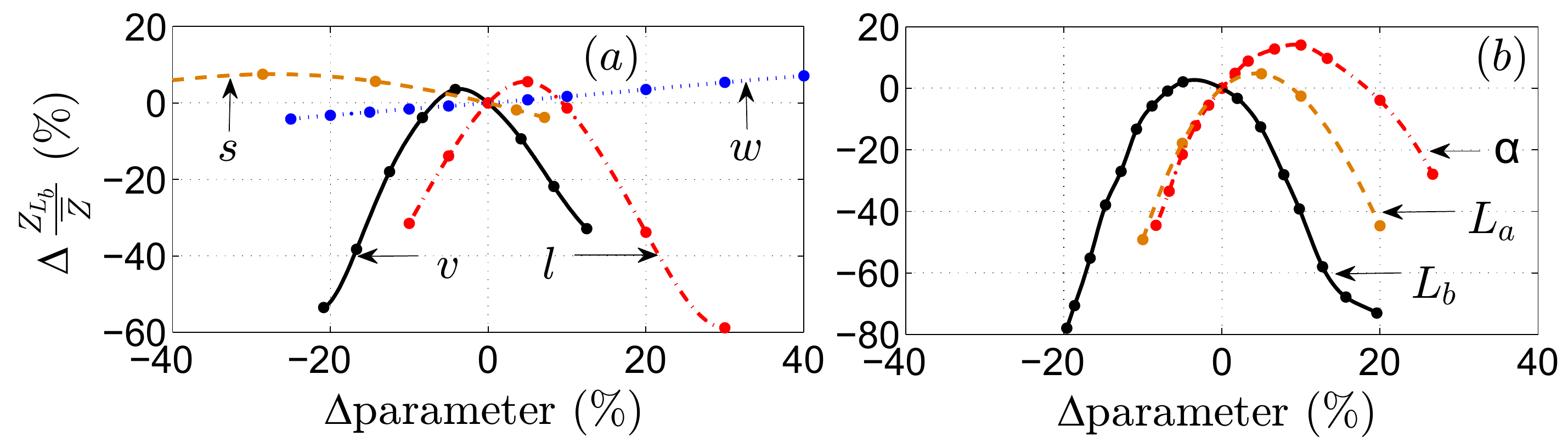}
\vspace{-0.in}
\caption[]{Sensitivity of the ratio of the maximum corner runup (normalized by the mean maximum runup) on $S_s$ $(Z_{L_b}/\overline Z)$ to changes in the landslide parameters ($s,w,l,v$, figure a) and lake's geometry parameters ($L_a,L_b,\alpha$, figure b). The changes in the parameters are given in percentage and are measured relative to the reference lake and landslide studied in \S4.2 (i.e. figures \ref{Snapshots}-\ref{Nonlin}). The other physical and simulation parameters are the same as in figure \ref{Snapshots}.}
\label{sensi}
\end{center}
\end{figure} 

In order to investigate the sensitivity of the presented results to the lake and landslide parameters, we study the effect of the lake's dimensions ($L_a,L_b,\alpha$) and slide's parameters ($s,w,l,v$) on the importance of the corner runup, measured as the ratio of maximum corner runup to the average runup on $S_s$, i.e. $Z_{L_b}/\overline Z$ in which $\overline Z=\f{1}{L_b}\int_0^{L_b}Z(a=0,b,c=0)~\d b$. Changes in the parameters and variables are measured relatively to the reference case studied in \S4.2 (i.e. figures \ref{Snapshots}-\ref{Nonlin}). The relative importance of the corner runup, displayed in figure \ref{sensi}, is clearly very sensitive to the lake's geometry (figure \ref{sensi}b). For example, a 20\% decrease in the lake's width leads to an 80\% drop in the relative corner runup. Figures \ref{sensi}b therefore further highlights that a change in the lake's geometry affects the location and time of positive and negative interference of edge- and outgoing-waves. The importance of the corner runup is also significantly altered by the length and the velocity of the landslide but is almost independent of the thickness and width of the slide (figure \ref{sensi}a). This suggests that uncertainties in the solid landslide's length or speed are likely to result in unreliable maximum wave runup predictions.

We would like to finally comment that long waves lose energy due to bottom friction which is caused by viscous effects near the seabed \cite[c.f.][]{Bernatskiy2012,Geist2009}. One way to model bottom friction is to introduce a friction term to the right-hand side of equations \eqref{927}. A number of different forms of such a dissipation term can be found in the literature that, generally, involve one or more free parameters that can be adjusted for model calibration \cite[e.g.][]{Satake1995,Synolakis2008,Zelt1990}. For example, the dissipation term suggested by \cite{Zelt1990} reads
\ba{fric}
-\f{K}{\mu}\f{(1+\f{\p (X^{(0)},Y^{(0)})}{\p (a,b)})}{h_0} \lp \begin{array}{cc} X_{t}^{(0)} \\ Y_{t}^{(0)} \end{array} \rp \sqrt{(X_{t}^{(0)})^2 + (Y_{t}^{(0)})^2}. 
\ea
With $K = 5\times10^{-5}$ and $\mu \tan\alpha = {3}/{60}$,  \cite{Zelt1990} showed that their numerical results compared well with the physical experiments of \cite{Synolakis1987} on nonbreaking solitary waves running up a 1/20 sloping beach. With the inclusion of this friction term into our model, we obtained less than 2\% maximum runup deviations from the runup values presented in figure \ref{nonlinab}. This indicates that dissipation by bottom friction can be safely neglected for the results presented here.

\section{Acknowledgements}
We would like to thank A. Zareei and B. S. Howard for careful reading of  the manuscript and helpful comments. L.A.C. was partially funded by the Jaehne Graduate Scholarship and the Frank and Margaret Lucas Scholarship. L.A.C. and M.R.A. gratefully acknowledge the support from the American Bureau of Shipping.

\section{Conclusions}
Under the long-wave and irrotationality assumptions, approximations of Airy type are applied to dynamical equations  in Lagrangian coordinates. 
A numerical scheme based on finite differences is employed to solve the Lagrangian equations and validated against existing theoretical/numerical results.  Aside from the convenience for  predicting  the motion of the shoreline, {which is of significant importance for nonlinear waves, we show that our model can predict reliable post-breaking 1D wave shapes when overturning occurs very close to the shoreline}. 

For a 2D tsunami in a lake generated by a landslide,   the combined influence of  edge-waves and radiated and reflected waves are examined for a rectangular-shaped basin.  It is found that nonlinearities can lead to a significant increase in  the maximum inundation  than  what is predicted by the linearized  equation, and that large runups occur near the lake's corners long after the submergence of the slide. {The contrast in runup between linear and nonlinear theories is shown to be a result of wave-seabed interactions as well as three-dimensional effects.} 

{The general features and physics reported here apply to the problem of landslide-generated tsunamis in a broad range of lake geometries and shapes. The details, however, may be very different. For instance, in a circular lake, edge-waves travel nonstop along the entire perimeter of the lake. This makes the subsequent interactions between edge-waves and outgoing-waves considerably different. Nevertheless, one would still expect that the combination of outgoing cross-lake waves with edge waves results in extreme inundations. Curved-boundary lakes (e.g. circular or elliptical) may in some cases experience further inundations enhanced by focusing of outgoing waves. Detailed investigation of such scenarios, while beyond the scope of the present manuscript, would be interesting and worth an independent study.}

\bibliographystyle{jfm}

\end{document}